\documentclass[10pt,twocolumn,letterpaper]{article}

\usepackage[pagenumbers]{cvpr}

\newcommand{\TODO}[1]{}

\makeatletter
\@namedef{ver@everyshi.sty}{2001/05/15 v3.00 EveryShipout Package (MS)}
\makeatother
\usepackage[accsupp]{axessibility}
\usepackage{amsmath,amssymb,amsfonts}
\usepackage{textcomp}
\usepackage{xcolor}
\usepackage{multirow}
\usepackage{booktabs}
\usepackage{microtype}
\usepackage{subcaption}
\usepackage{diagbox}
\usepackage{colortbl}
\definecolor{rowcolor}{rgb}{0.85, 0.95, 0.9}
\newenvironment{tcolorbox}[1][]{
  \def\tcbempty{}\def\tcbarg{#1}
  \par\medskip\noindent\rule{\linewidth}{0.8pt}\par\smallskip
  \ifx\tcbarg\tcbempty\else
    \noindent{\small\bfseries\tcbparsetitle#1,}\par\smallskip
    \noindent\rule{\linewidth}{0.4pt}\par\smallskip
  \fi
  \small
}{
  \par\smallskip\noindent\rule{\linewidth}{0.8pt}\par\medskip
}
\def\tcbparsetitle title=#1,{#1}
\usepackage{float}
\usepackage{makecell}
\usepackage{tikz}
\usetikzlibrary{shapes.geometric, arrows.meta, positioning, fit, backgrounds,trees}
\usepackage{scalerel}
\usepackage{arydshln}

\newcommand{\mathbbm}[1]{\text{\usefont{U}{bbm}{m}{n}#1}}
\DeclareCaptionStyle{ruled}{labelfont=normalfont,labelsep=colon,strut=off}

\definecolor{cvprblue}{rgb}{0.21,0.49,0.74}
\usepackage[breaklinks,colorlinks,allcolors=cvprblue]{hyperref}

\def\paperID{*****}
\def\confName{CVPR}
\def\confYear{2026}

\title{HippoMM: Hippocampal-inspired Multimodal Memory\\ for Long Audiovisual Event Understanding}

\author{Yueqian Lin$^{1}$ \quad Jingyang Zhang$^{2}$ \quad Qinsi Wang$^{1}$ \quad Hancheng Ye$^{1}$\\
Yuzhe Fu$^{1}$ \quad Yudong Liu$^{1}$ \quad Hai Helen Li$^{1}$ \quad Yiran Chen$^{1}$\\
$^{1}$Duke University \quad $^{2}$Independent Researcher\\
{\tt\small yueqian.lin@duke.edu}
}

\begin{document}
\maketitle
\begin{abstract}
Comprehending extended audiovisual experiences remains challenging for computational systems, particularly temporal integration and cross-modal associations fundamental to human episodic memory.
We introduce \textbf{HippoMM}, a \textbf{computational cognitive architecture} that maps hippocampal mechanisms to solve these challenges.
Rather than relying on scaling or architectural sophistication, HippoMM implements three integrated components:
(i) \textbf{Episodic Segmentation} detects audiovisual input changes to split videos into discrete episodes, mirroring dentate gyrus pattern separation; (ii) \textbf{Memory Consolidation} compresses episodes into summaries with key features preserved, analogous to hippocampal memory formation; and (iii) \textbf{Hierarchical Memory Retrieval} first searches semantic summaries, then escalates via temporal window expansion around seed segments for cross-modal queries, mimicking CA3 pattern completion. These components jointly create an integrated system exceeding the sum of its parts. On our HippoVlog benchmark testing associative memory, HippoMM achieves state-of-the-art \textbf{78.2\%} accuracy while operating \textbf{5x} faster than retrieval-augmented baselines. Our results demonstrate that cognitive architectures provide blueprints for next-generation multimodal understanding.
\end{abstract}
\section{Introduction}
\label{sec:introduction}

Despite recent progress, current multimodal models still cannot perform long-form audiovisual understanding accurately and efficiently.
Specifically, they (i) must retain entire continuous content spanning several hours as the context \cite{zhang2024longva}, (ii) cannot reconstruct complete experiences when given partial sensory cues \cite{lei2021less}, and (iii) cannot form abstract memories that transfer across contexts \cite{wang2022internvideo}.
Such failures prevent machines from seamlessly understanding real-world experiences the way humans naturally do: we effortlessly (i) segment raw, continuous content into meaningful episodes \cite{zacks2007event}, (ii) recall entire episodes from a single sound or image \cite{tulving1983elements}, and (iii) extract lasting knowledge from fleeting perceptions~\cite{mcclelland1995complementary}.

\begin{figure}[ht]
    \centering
    \includegraphics[width=\linewidth]{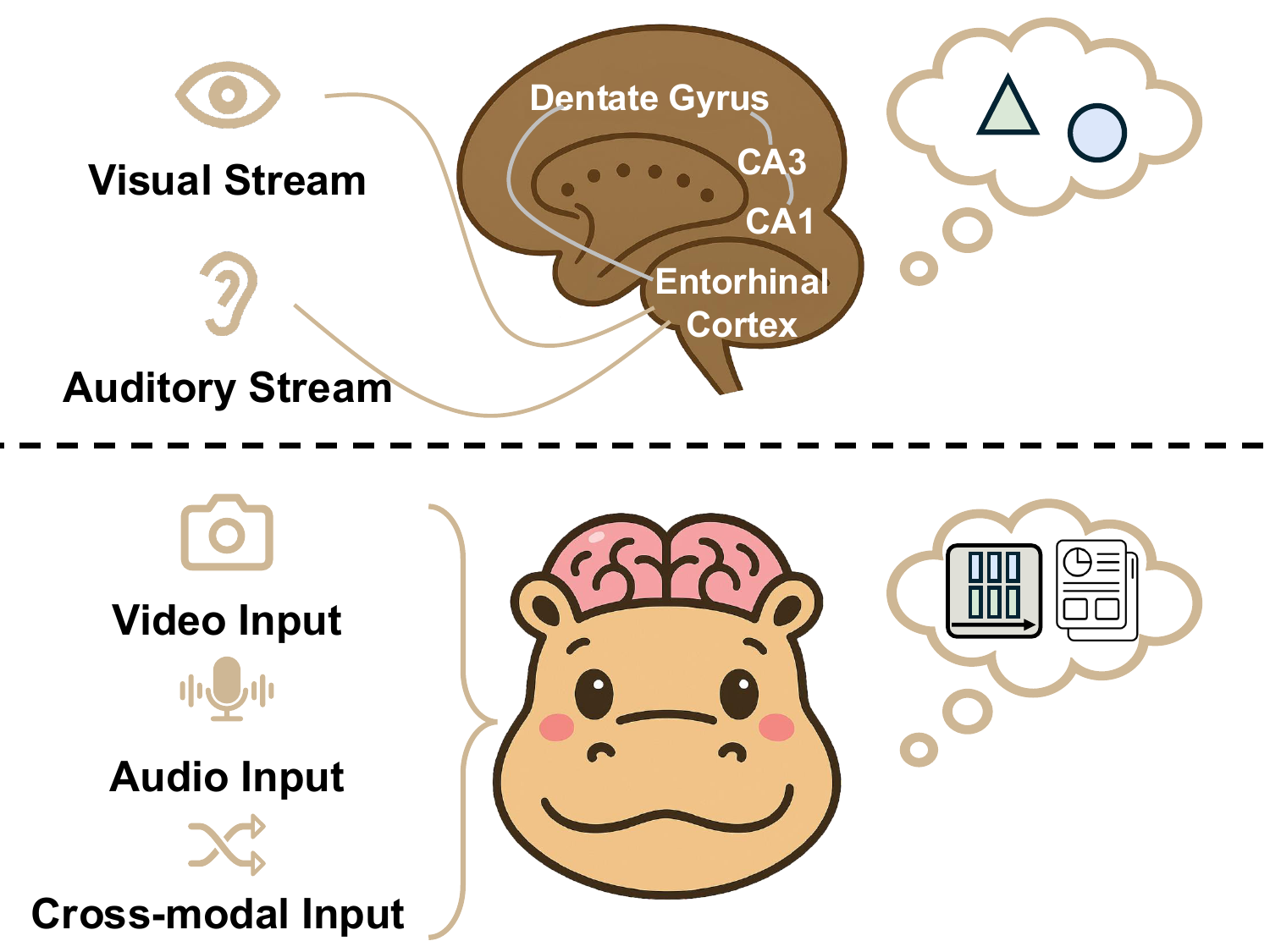}
    \caption{\textbf{Conceptual Overview of Hippocampal vs. HippoMM Processing.} The biological hippocampus (top) and our HippoMM architecture (bottom) both integrate continuous sensory streams to form and recall episodic memories.}
    \label{fig:intro}
\end{figure}

While prior works improve multimodal systems by scaling up or introducing sophisticated architectures~\cite{radford2021learning, girdhar2023imagebind}, recent memory-augmented approaches~\cite{park2024malmm} show explicit memory mechanisms are crucial. We posit that current systems lack effective cognitive mechanisms for episodic memory.
The biological hippocampus provides a cognitive architecture that solves these challenges through three core mechanisms: segmentation of continuous experience into discrete episodes, associative reconstruction from fragmentary inputs, and consolidation of perceptual details into semantic knowledge~\cite{yassa2011pattern, rolls2013quantitative}.
Figure~\ref{fig:intro} illustrates the parallel between biological hippocampal processing and our computational approach.
By translating these principles into algorithmic implementations, we design an accurate and efficient multimodal system.

To this end, in this work we propose \textbf{HippoMM}, which implements our insight through direct functional mapping of hippocampal computations.
Following dentate gyrus (DG) pattern separation \cite{yassa2011pattern}, our \textit{Episodic Segmentation} employs content-adaptive boundaries to parse inputs into episodic codes.
Emulating hippocampal consolidation processes \cite{kumaran2016learning}, \textit{Memory Consolidation} progressively transforms perceptual traces into semantic representations, culminating in ThetaEvent memories.
Mimicking CA3 autoassociation~\cite{rolls2013quantitative}, \textit{Hierarchical Memory Retrieval} enables pattern completion from partial cues and cross-modal queries.
This integrated architecture amplifies each component through their interactions, achieving performance gains that validate the cognitive architectural approach.

Another contribution of our work is \textbf{HippoVlog}, a benchmark designed to test cross-modal associative recall in long-form video understanding.
Unlike existing benchmarks that focus on sequential comprehension \cite{mangalam2023egoschema} or simple retrieval tasks \cite{yu2019activitynetqa}, HippoVlog evaluates episodic memory functions through partial cue reconstruction and semantic consolidation questions.
Built from 25 daily vlogs, it provides 1,000 manually-validated questions across four memory function categories.

In summary, our contributions are:
\begin{itemize}
    \item We design \textbf{HippoMM}, which implements hippocampal memory functions through \textbf{episodic segmentation} at perceptual boundaries, \textbf{cross-modal feature binding} in consolidated memory units, and \textbf{hierarchical retrieval} that balances semantic summaries with detailed recall.
    \item We introduce \textbf{HippoVlog}, a benchmark for \textbf{cross-modal associative memory} with questions requiring pattern completion from partial cues, addressing the gap in existing benchmarks that only test sequential comprehension.
    \item We demonstrate that \textbf{cognitive architectures provide principled solutions}: HippoMM achieves \textbf{78.2\%} accuracy and \textbf{5×} speedup over retrieval baselines, validating biological mechanisms as effective computational designs.
\end{itemize}

\section{Related Work}
\label{sec:related_work}

\subsection{Hippocampal Memory Mechanisms}
The hippocampus performs three key functions: \textit{pattern separation} through sparse coding in the DG \cite{yassa2011pattern}, \textit{pattern completion} via CA3 autoassociative networks~\cite{rolls2013quantitative}, and event segmentation at boundaries~\cite{baldassano2017discovering}. While computational models have adapted these principles \cite{kumaran2016learning}, they primarily focus on unimodal contexts, overlooking the hippocampus's role as a \textbf{multimodal integrator} \cite{eichenbaum2014time}.

\subsection{AudioVisual Multimodal Understanding}
Current multimodal systems employ joint embeddings \cite{radford2021learning, girdhar2023imagebind} or video-language models \cite{alayrac2022flamingo,liu2025llavida} but lack mechanisms for temporal memory organization and associative recall. While recent models like VideoLLaMA 2 \cite{cheng2024videollama2} and Qwen2.5-Omni \cite{xu2025qwen2} incorporate audiovisual processing, and voice-native benchmarks \cite{lin2026vera} expose modality-induced reasoning gaps, these efforts focus on modality alignment rather than memory formation. Related benchmarks such as LongVALE~\cite{geng2025longvale} target temporal grounding and dense captioning—predicting temporal boundaries in pre-segmented video—which is complementary to but distinct from our task of cross-modal associative recall from continuous streams. Critically, existing approaches cannot form episodic memories from continuous streams or perform cross-modal pattern completion, where sounds trigger visual memories or vice versa. This fundamental gap in hippocampal-like memory mechanisms represents a major limitation, as without them, models process only pre-segmented clips through attention \cite{tsai2019multimodal} or fusion \cite{nagrani2021attention,lin2025speechprune,wang2025corematching} without true associative recall.

\subsection{Biomimetic Memory Architectures}
Memory-augmented networks like the Differentiable Neural Computer \cite{graves2016hybrid} pioneered external memory but remain predominantly unimodal. Recent hippocampal-inspired architectures include HippoRAG \cite{gutierrez2024hipporag} for text retrieval and MA-LMM \cite{park2024malmm} which extends memory-augmented approaches to long-term video understanding through memory banks. HippoMM uniquely integrates pattern separation, consolidation, and cross-modal pattern completion in a unified architecture for continuous audiovisual understanding.

\begin{figure*}[ht]
    \centering
    \includegraphics[width=1\linewidth]{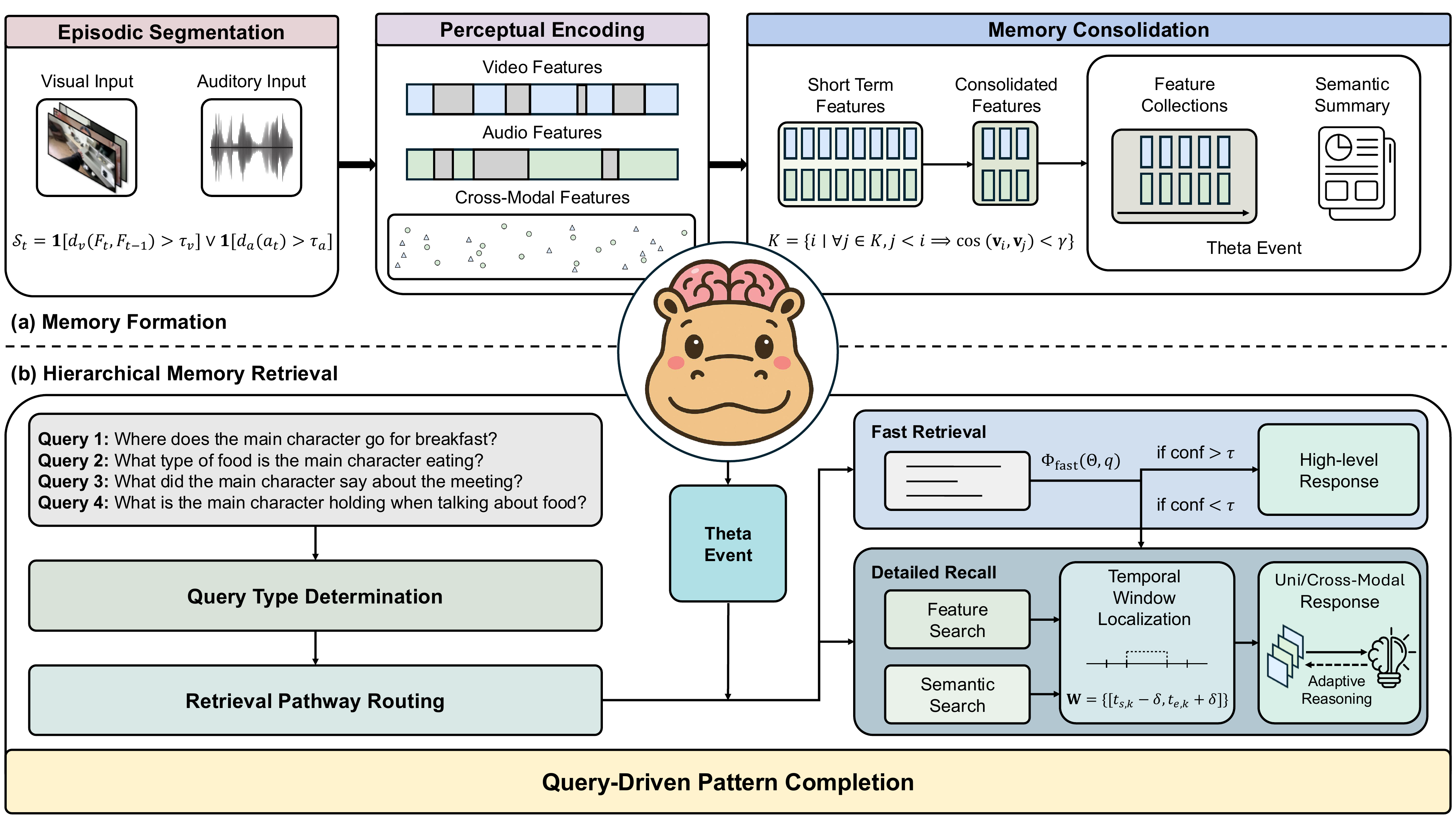}
    \caption{The HippoMM architecture. (a) Memory Formation: composed of Episodic Segmentation ($\mathcal{S}_t$) based on perceptual boundaries, Perceptual Encoding of visual and auditory inputs with cross-modal features, and Memory Consolidation that progressively transforms short-term features through similarity-based filtering ($K$) into consolidated features, feature collections, and ultimately semantic summaries to form ThetaEvent ($\theta$). (b) Hierarchical Memory Retrieval: implementing Query-Driven Pattern Completion through fast retrieval ($\Phi_{\text{fast}}$) and detailed recall pathways with temporal window localization ($\mathbf{W}$).}
\end{figure*}

\section{Multimodal Memory-augmented Retrieval}
\label{sec:mmr}

\subsection{Task Formulation}
We formalize the task of Multimodal Memory-augmented Retrieval (MMR). While retrieval tasks have been extensively studied across various modalities, existing formulations typically treat retrieval as matching queries against pre-indexed content. Our work uniquely addresses retrieval from continuous multimodal streams requiring active memory formation, which is fundamentally different from traditional settings as the system must continuously construct and maintain structured memory representations from ongoing inputs, mimicking human episodic memory.

The objective is to develop a system that processes a long-form, continuous multimodal data stream $X = \{x_t\}$ and, given a query $q$, synthesizes an accurate answer $a$ through three functional stages: (i) memory formation $\mathcal{F}_{\text{mem}}$ transforms $X$ into structured representation $M$; (ii) retrieval $\mathcal{R}_{\text{retrieval}}$ identifies relevant evidence $E$ from $M$; and (iii) answer synthesis $\mathcal{G}_{\text{synth}}$ generates response $a$. HippoMM instantiates $\mathcal{F}_{\text{mem}}$ via content-adaptive segmentation, cross-modal encoding, and similarity-filtered consolidation; $\mathcal{R}_{\text{retrieval}}$ via GPT-4o-based confidence-gated hierarchical retrieval; and $\mathcal{G}_{\text{synth}}$ via LLM-based adaptive reasoning—each grounded in hippocampal principles but not the only valid instantiation.

\subsection{The HippoVlog Benchmark}
A significant challenge in evaluating MMR systems is the lack of benchmarks designed to test for cross-modal associative recall. Recent benchmarks have advanced long-form video understanding significantly: MLVU \cite{zhou2025mlvu} evaluates multi-task comprehension across diverse genres, ALLVB \cite{zhang2025allvb} unifies nine tasks with 2-hour videos, and Video-MME \cite{chen2025videomme} provides comprehensive multimodal evaluation. However, these benchmarks fundamentally test \textit{comprehension} of presented information rather than \textit{memory formation and associative recall}.

Specifically, existing benchmarks evaluate whether models can answer questions about explicitly shown content, but they do not test: (i) the ability to form episodic memories from continuous experience that enable later reconstruction from partial cues, (ii) cross-modal pattern completion where a sound triggers recall of associated visual content or vice versa, and (iii) the consolidation of perceptual details into transferable semantic knowledge. These capabilities, central to human episodic memory, require fundamentally different evaluation paradigms than sequential comprehension.

To address this gap, we introduce \textbf{HippoVlog}, a benchmark specifically designed to evaluate hippocampal-like memory functions.
Built from 25 long-form daily vlogs totaling 682 minutes, it provides 1,000 questions across four memory function categories: Cross-modal binding ($T_{V \times A}$), Auditory-focused retrieval ($T_A$), Visual-focused retrieval ($T_V$), and Semantic reasoning ($T_S$).
The quality of the benchmark is ensured through rigorous manual validation, achieving a high inter-annotator agreement (Cohen's $\kappa=0.975$).
Unlike comprehension-focused benchmarks, HippoVlog's questions probe whether systems can reconstruct complete multimodal experiences from fragmentary cues—the hallmark of episodic memory.
Construction details are provided in the Appendix. Our code is publicly available.\footnote{Code: \url{https://github.com/linyueqian/HippoMM}}

\section{Methodology}
\label{sec:methodology}

We implement the fundamental cognitive architecture of the hippocampus, translating its three core computations of pattern separation (DG), pattern completion (CA3), and memory consolidation (CA1) into algorithms for multimodal understanding.
This hippocampal-inspired architecture addresses precisely those challenges that confound current models: segmenting continuous experience into discrete episodes, reconstructing complete memories from partial cues, and abstracting semantic knowledge from perceptual details.
Our methodology implements these biological principles through a two-phase process: a \textbf{Memory Formation} phase that mirrors the hippocampal encoding pathway from entorhinal cortex through DG-CA3-CA1, and a \textbf{Hierarchical Memory Retrieval} phase that implements CA3 autoassociation and CA1 semantic recall for query-driven reasoning.

\subsection{Memory Formation}
\label{subsec:memory-formation}
Our memory formation pipeline implements three computational primitives from the hippocampus: pattern separation (episodic segmentation), perceptual encoding, and memory consolidation (semantic abstraction).
These stages transform continuous multimodal streams into a hierarchical memory structure with both detailed episodic traces and efficient semantic indices.

\noindent\textbf{Episodic Segmentation} \textit{(Pattern Separation / Threshold-based Boundary Detection).}
Following the DG's pattern separation function that transforms overlapping inputs into distinct neural codes, we implement content-adaptive temporal segmentation to prevent interference between similar memories.
Fixed-window chunking arbitrarily divides videos at regular intervals, often splitting coherent events mid-action or grouping unrelated scenes together.
To preserve semantic integrity, our architecture segments the stream by detecting perceptual discontinuities that align with natural event boundaries.
Specifically, we trigger a boundary at time $t$ when either visual similarity drops significantly between consecutive frames or level drops below a threshold indicating silence or scene breaks:
\begin{equation}
    \mathcal{S}_t = \mathbbm{1}[d_v(F_t, F_{t-1}) > \tau_v] \vee \mathbbm{1}[d_a(a_t) > \tau_a],
    \label{eq:segmentation_trigger}
\end{equation}
where visual dissimilarity $d_v$ is measured using Structural Similarity Index Measure (SSIM) (Eq.~\ref{eq:ssim_distance}) with threshold $\tau_v$ implementing the sparsity constraint observed in DG neurons, and audio boundary detection $d_a$ (Eq.~\ref{eq:audio_energy}) uses threshold $\tau_a$ chosen through empirical validation to identify pauses in speech or sound that often coincide with activity transitions.
\begin{align}
    d_v(F_t, F_{t-1}) &= 1 - \text{SSIM}(F_t, F_{t-1}) \label{eq:ssim_distance}, \\
    d_a(a_t) &= -20\log_{10}\left(\sqrt{\frac{1}{N}\sum_{i=1}^{N}a_i^2}\right).\label{eq:audio_energy}
\end{align}
These perceptually-motivated boundaries produce segments constrained between $t_{min} = 5$ and $t_{max} = 10$ seconds, aligning with human event segmentation timescales reported in prior work~\cite{zacks2007event}.
This content-aware segmentation produces a series of coherent temporal windows, $[t_{s,i}, t_{e,i}]$, forming the foundational units for memory encoding.

\noindent\textbf{Perceptual Encoding.}
Mirroring how the entorhinal cortex provides multimodal input transformed through DG into CA3's distributed representations, we employ specialized encoders that preserve both modality-specific features and cross-modal associations.
Within each event segment, the system captures a rich, multimodal representation of the experience.
This is not a monolithic encoding but a tripartite strategy that orchestrates specialized foundation models: a vision-language model generates dense textual descriptions ($\mathcal{T}_v$), a speech recognition model provides precise transcriptions ($\mathcal{T}_a$), and a cross-modal embedding model ($\mathcal{E}_c$) maps both visual frames and audio snippets into a shared feature space, enabling direct similarity comparisons across modalities.
This distributed encoding mirrors CA3's recurrent architecture where diverse inputs converge into unified episodic representations.
These outputs are aggregated into a detailed episodic record, the \texttt{ShortTermMemory} object $m_i$:
\begin{equation}
    m_i = \{\mathbf{E}_i, \mathbf{T}_i, \mathbf{C}_i, t_{s,i}, t_{e,i}\},
    \label{eq:short_term_memory}
\end{equation}
where $\mathbf{E}_i$ denotes the sequence of cross-modal embeddings from $\mathcal{E}_c$, $\mathbf{T}_i$ aggregates the textual outputs from both $\mathcal{T}_v$ (visual descriptions) and $\mathcal{T}_a$ (audio transcriptions), $\mathbf{C}_i$ contains pointers to the raw audiovisual data, and [$t_{s,i}, t_{e,i}$] marks the temporal boundaries of segment $i$.

\noindent\textbf{Memory Consolidation} \textit{(Synaptic Consolidation / Cosine-Similarity Sparse Filtering).}
Implementing synaptic and systems consolidation (with CA3/CA1 roles), our pipeline progressively abstracts from perceptual details to semantic summaries.
Storing every fine-grained detail from a long stream is inefficient and leads to catastrophic interference between similar events.
Our architecture addresses this with a consolidation mechanism that actively filters for novelty.
This process filters redundant \texttt{ShortTermMemory} objects, retaining only those that introduce significant new information.
For each segment $m_i$, a representative embedding $\mathbf{v}_i$ is computed by averaging the sequence of cross-modal embeddings $\mathbf{E}_i$ produced by the embedding model $\mathcal{E}_c$.
The segment is retained in the consolidated set $K$ only if its embedding is sufficiently dissimilar from all previously stored memories, governed by a similarity threshold $\gamma$:
\begin{equation}
    K = \{i \mid \forall j \in K, j<i \implies \cos(\mathbf{v}_i, \mathbf{v}_j) < \gamma \}.
    \label{eq:consolidation_filtering}
\end{equation}
This threshold $\gamma$ is chosen to encourage sparse retention, analogous to CA3 sparsity where only 2-5\% of neurons are active for any given memory.
This selective filtering creates an efficient memory store that is robust to distraction and optimized for retrieval.
The consolidated set $K$ contains indices of retained segments. For clarity, we use $k \in K$ to denote consolidated segment indices and $m_k$ for their corresponding memory objects.

The cornerstone of our architecture's efficiency is the creation of a high-level semantic index over the consolidated episodic records.
This stage transforms detailed, fine-grained memories into abstract, language-based summaries, making vast stretches of time searchable at near-zero cost.
For each consolidated memory $m_k$ where $k \in K$ denotes indices from the filtered set, an LLM ($\phi_{\text{LLM}}$) synthesizes its multimodal contents into a concise textual ``gist,'' $\mathbf{S}_{\theta_k}$:
\begin{equation}
    \mathbf{S}_{\theta_k} = \phi_{\text{LLM}}(\text{FormattedContext}(m_k)),
    \label{eq:semantic_summary_revised}
\end{equation}
where $\text{FormattedContext}(\cdot)$ structures the multimodal content from $m_k$ (visual descriptions, transcriptions, and temporal information) into a prompt format suitable for the LLM.
This summary becomes the centerpiece of a \texttt{ThetaEvent} object, our long-term memory data structure.
The name ``ThetaEvent'' reflects hippocampal theta oscillations that bind experiences into coherent episodes.
The \texttt{ThetaEvent} bundles the abstract summary with the representative embedding $\mathbf{v}_k$ and a pointer back to its detailed source memory, creating a powerful dual-representation system that bridges semantic abstraction and perceptual detail, precisely the function performed by CA1 in biological memory consolidation.
\begin{equation}
    \theta_k = \{\mathbf{v}_k, \mathbf{S}_{\theta_k}, \mathbf{T}_{\theta_k}, \mathbf{I}_{\theta_k}, \mathbf{A}_{\theta_k}\},
    \label{eq:theta_event}
\end{equation}
where $\mathbf{v}_k$ is the representative embedding, $\mathbf{S}_{\theta_k}$ is the generated semantic summary, $\mathbf{T}_{\theta_k}$ contains temporal references linking to the source $m_k$, and $\mathbf{I}_{\theta_k}$, $\mathbf{A}_{\theta_k}$ store key visual and auditory context used in summary generation.

\subsection{Hierarchical Memory Retrieval \textit{(Pattern Completion / Confidence-Gated Two-Path Lookup)}}
\label{subsec:memory-retrieval}
The hippocampus supports two complementary retrieval modes: rapid semantic access through CA1 and detailed pattern completion through CA3 recurrence.
Our retrieval architecture directly implements this biological duality, resolving the critical trade-off between speed and fidelity by dynamically choosing its search strategy based on the query's demands, facilitating complex forms of reasoning such as associative recall and cross-modal pattern completion.

\noindent\textbf{Query-Driven Pathway Selection.} Implementing the hippocampus's dual retrieval modes: CA1's rapid semantic access and CA3's detailed pattern completion, we employ confidence-gated hierarchical retrieval.
Upon receiving a query $q$, the system first attempts \textbf{Fast Retrieval} ($\Phi_{\text{fast}}$), implementing CA1's semantic pathway by searching exclusively on the abstract \texttt{ThetaEvent} summaries.
This path rapidly resolves high-level and gist-based queries.
Only if the confidence of this initial result falls below a tuned threshold $\tau$ does the system escalate to the more resource-intensive \textbf{Detailed Recall} ($\Psi_{\text{detailed}}$), which engages CA3's pattern completion mechanisms.
This confidence score is generated by the LLM's internal calibration when attempting to answer from summaries alone, reflecting its assessment of whether sufficient information exists in the abstract representations.
This on-demand escalation provides a ``zoom-in'' capability, accessing the rich, consolidated \texttt{ShortTermMemory} objects for evidence-grounded reasoning.
\begin{equation}
    \mathcal{R}(q) =
    \begin{cases}
      \Phi_{\text{fast}}(\Theta, q) & \text{if } \text{conf}(\Phi_{\text{fast}}) > \tau, \\
      \Psi_{\text{detailed}}(q, \mathcal{M}, \mathcal{Q}_{\text{type}}(q)) & \text{otherwise},
    \end{cases}
    \label{eq:hierarchical_retrieval}
\end{equation}
where $\Theta = \{\theta_k \mid k \in K\}$ denotes the collection of \texttt{ThetaEvent} objects.

\noindent\textbf{Detailed Recall with Cross-Modal Pattern Completion.}
CA3's recurrent connectivity enables pattern completion: reconstructing complete memories from partial cues.
This biological computation is most powerful for cross-modal binding, where a cue in one modality triggers recall of associated information in another modality.
When detailed recall is invoked, the system deploys a specialized search strategy tailored to the query's modality.
For single-modality queries, it performs either an embedding-based \textit{Feature Search} or a text-based \textit{Semantic Search}.
The key innovation, however, is its mechanism for cross-modal associative recall, which uses temporal co-occurrence as a powerful proxy for episodic binding, mirroring how CA3 neurons bind coincident inputs through synaptic plasticity.

This process begins by using the query cue to find relevant seed segments in the target memory store:
\begin{equation}
    \mathbf{S}_{\text{query}} = \text{TopK}(\text{sim}(q_{embed}, \{\mathbf{v}_k \mid m_k \in \mathcal{M}\}), k). \label{eq:cross_modal_query_retrieve_revised}
\end{equation}
It then defines expanded temporal windows $\mathbf{W}$ around these seeds to capture the surrounding event context, implementing the temporal receptive fields of hippocampal place cells.
\begin{equation}
    \mathbf{W} = \{[t_{s,k} - \delta, t_{e,k} + \delta] \mid k \in \mathbf{S}_{\text{query}} \}. \label{eq:cross_modal_window}
\end{equation}
Finally, it retrieves information from the target modality by searching for any memory segments that temporally overlap with these windows, effectively reconstructing the integrated experience.
\begin{multline}
\mathbf{S}_{\text{target}} = \{\text{Extract}_{\text{target}}(m_j) \mid \\
    (t_{s,j}, t_{e,j}) \text{ overlaps with } \mathbf{W}\},
    \label{eq:cross_modal_target_retrieve}
\end{multline}
where $\text{Extract}_{\text{target}}(m_j)$ extracts modality-specific content from segment $m_j$: visual descriptions from $\mathbf{T}_j$ for visual targets, or transcriptions for auditory targets. We denote segment $j$'s temporal boundaries as $(t_{s,j}, t_{e,j})$.

To illustrate cross-modal pattern completion, consider the query: ``What was on the screen when the applause started?''
The system identifies the auditory cue (``applause'') and visual target (``on the screen''), with low confidence from Fast Retrieval triggering Detailed Recall due to the specific temporal link.
Following Eq.~\ref{eq:cross_modal_query_retrieve_revised}, it finds seed segment $m_k$ containing the applause sound, then forms temporal window $\mathbf{W}$ around it (Eq.~\ref{eq:cross_modal_window})—implementing how hippocampal neurons encode temporally extended episodes.
Visual information from segments overlapping $\mathbf{W}$ is retrieved (Eq.~\ref{eq:cross_modal_target_retrieve}) and passed to Adaptive Reasoning for synthesis.
\begin{table*}[t]
\caption{\textbf{Performance comparison of HippoMM against existing methods and ablation variants on HippoVlog.} PT: Processing Time; ART: Average Response Time; A+V: Cross-modal (Audio+Visual) accuracy; A: Audio-only accuracy; V: Visual-only accuracy; S: Semantic understanding accuracy. Best results are in \textbf{bold}, second best are \underline{underlined}. HippoMM significantly outperforms prior methods in all modality-specific tasks while maintaining efficient processing and response times. Ablation studies demonstrate the importance of each component: Detailed Recall (DR), Fast Retrieval (FR), and Adaptive Reasoning (AR).}
\centering
\begin{tabular}{l c c cccc c}
\toprule
\multirow{2}{*}{\textbf{Method}} & \multirow{2}{*}{\textbf{PT $\downarrow$}} & \multirow{2}{*}{\textbf{ART $\downarrow$}} & \multicolumn{4}{c}{\textbf{Modality Performance}} & \multirow{2}{*}{\textbf{Avg. Acc. $\uparrow$}} \\
\cmidrule(lr){4-7}
 & & & \textbf{A+V $\uparrow$} & \textbf{A $\uparrow$} & \textbf{V $\uparrow$} & \textbf{S $\uparrow$} & \\
\midrule
\multicolumn{8}{l}{\textit{\textbf{Prior Methods}}} \\
NotebookLM~\citep{google2024notebooklm} & -- & -- & 28.40\% & 23.20\% & 28.00\% & 26.80\% & 26.60\% \\
VideoRAG~\citep{ren2024videorag} & 9.46h & 112.5s & 63.6\% & 67.2\% & 41.2\% & 84.8\% & 64.2\% \\
Ola~\citep{liu2025ola} & -- & 79.4s & 72.4\% & 85.6\% & 57.6\% & 84.0\% & 74.9\% \\
VideoLLaMA 3~\citep{zhang2025videollama3} & -- & 58.3s & - & - & 70.8\% & 75.2\% & 73.0\% \\
Simple-Assoc-Memory (SAM) & -- & -- & 32.0\% & 32.8\% & 33.2\% & 23.2\% & 30.3\% \\
Gemini-3-Flash~\citep{google2025gemini3flash} & -- & -- & 44.0\% & 45.2\% & 40.0\% & 56.0\% & 46.3\% \\
GPT-5~\citep{openai2025gpt5} & -- & -- & \textbf{72.0\%} & 73.2\% & 45.6\% & 88.0\% & 69.7\% \\
Qwen2.5-Omni~\citep{xu2025qwen2} & -- & -- & 71.2\% & 69.6\% & 51.2\% & 73.6\% & 66.4\% \\
\midrule
\multicolumn{8}{l}{\textit{\textbf{Ablation Studies}}} \\
HippoMM w/o DR, AR & 5.09h & \textbf{4.14s} & 66.8\% & 73.2\% & 60.4\% & 90.0\% & 72.6\% \\
HippoMM w/o FR, AR & 5.09h & 27.3s & \textbf{72.0\%} & 80.0\% & \textbf{66.8\%} & 83.2\% & 75.5\% \\
HippoMM w/o AR & 5.09h & \underline{11.2s} & 68.8\% & \underline{80.8\%} & \underline{65.6\%} & \underline{92.0\%} & \underline{76.8\%} \\
HippoMM w/o DR & 5.09h & 6.39s & 39.2\% & 65.6\% & 48.0\% & 92.0\% & 61.2\% \\
HippoMM w/o FR & 5.09h & 19.54s & 72.0\% & 79.6\% & 63.6\% & 83.2\% & 74.6\% \\
HippoMM (Qwen2.5-14B) & 5.09h & 15.7s & 69.3\% & 77.1\% & 56.6\% & 76.4\% & 70.8\% \\
HippoMM (EOR-only) & 5.09h & -- & 63.2\% & 70.4\% & 60.8\% & 90.0\% & 71.1\% \\
HippoMM ($\gamma=0.80$) & 5.09h & 20.4s & 69.2\% & 78.4\% & 64.0\% & 90.8\% & 75.6\% \\
HippoMM ($\gamma=0.90$) & 5.09h & 20.4s & 71.6\% & 78.8\% & 64.4\% & 90.0\% & 76.2\% \\
\midrule
\rowcolor{blue!20} \textbf{HippoMM (Ours)} & \textbf{5.09h} & 20.4s & \underline{70.8\%} & \textbf{81.6\%} & \textbf{66.8\%} & \textbf{93.6\%} & \textbf{78.2\%} \\
\bottomrule
\end{tabular}
\vspace{-1em}
\label{tab:method_comparison}
\end{table*}

\noindent\textbf{Adaptive Reasoning.}
While the hippocampus provides memory access, the prefrontal cortex performs executive control and conflict resolution.
Our Adaptive Reasoning module implements this cognitive function, particularly crucial when reconciling outputs from the dual retrieval pathways.
The module $\Gamma$ serves as a locus for cognitive reflection, especially when the system escalates from Fast Retrieval to Detailed Recall.
In such cases, $\Gamma$ receives both summary-based evidence from $\Phi_{\text{fast}}$ and fine-grained evidence from $\Psi_{\text{detailed}}$, reconciling potentially conflicting accounts.
For instance, if a summary suggests general activity while detailed recall reveals specific contradictory actions, the module prioritizes high-fidelity, grounded evidence from the detailed pathway—mirroring prefrontal-hippocampal interactions where top-down control resolves memory conflicts.
The module synthesizes a final answer that accurately represents the most reliable evidence:
\begin{equation}
    a = \mathcal{R}_{\text{final}}(q) = \Gamma(q, r_{\text{retrieved}}, \mathcal{E}_{\text{context}}).
    \label{eq:adaptive_reasoning}
\end{equation}
By integrating this reflective step, the architecture ensures its output is not just retrieved facts but a reasoned interpretation, demonstrating that cognitive architectures provide both biological plausibility and functional superiority.

\section{Experiments}
\label{sec:experiments}

\subsection{Implementation Details}

Our system, HippoMM, was implemented on an NVIDIA L40S GPU.
Episodic segmentation processed video frames at 1-10 fps using SSIM and audio energy thresholds to create 5-10s segments.
Perceptual Encoding generated \texttt{ShortTermMemory} objects ($m_i$, Eq.~\ref{eq:short_term_memory}) containing 1024-dim ImageBind embeddings~\citep{girdhar2023imagebind} ($\mathcal{E}_c$), Whisper transcriptions~\citep{radford2023robust} ($\mathcal{T}_a$), Qwen2.5-VL visual descriptions~\citep{Qwen2.5-VL} ($\mathcal{T}_v$), and raw data pointers ($\mathbf{C}_i$).
Memory Consolidation filtered segments based on cosine similarity ($\gamma=0.85$) between average ImageBind embeddings.
Qwen2.5-VL~\citep{Qwen2.5-VL} generated textual summaries ($\mathbf{S}_{\theta_k}$) for \texttt{ThetaEvent} objects.
For Hierarchical Memory Retrieval, GPT-4o~\citep{hurst2024gpt} served as query classifier and final answer synthesizer ($\Gamma$), while Qwen2.5-VL provided confidence scoring during Fast Retrieval ($\tau=0.75$ threshold) and fine-grained visual analysis during Detailed Recall.
\textit{For complete configurations and prompt templates, see Appendix.}
\begin{figure*}[t]
      \centering
      \includegraphics[width=\linewidth]{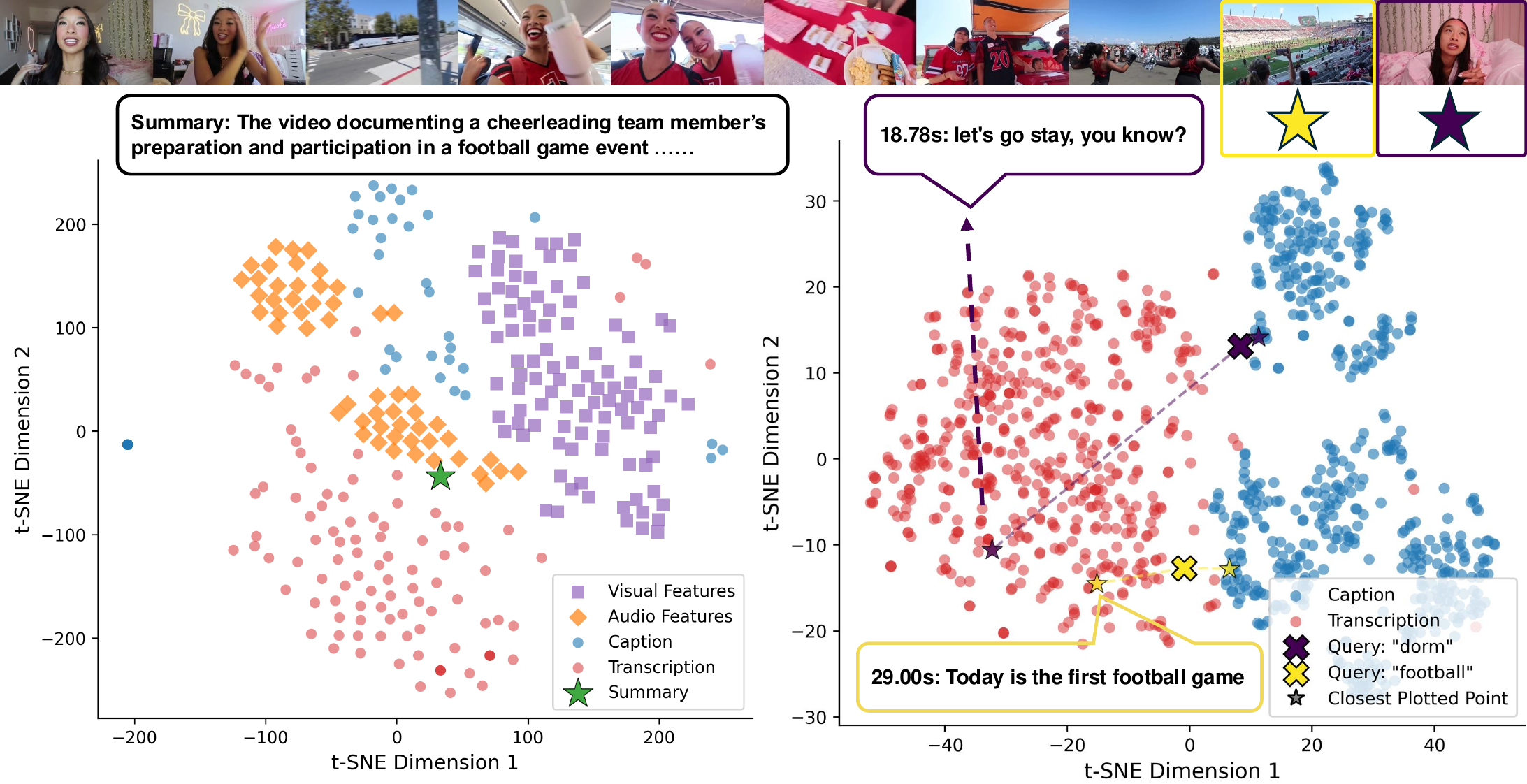}
      \caption{Visualization of consolidated \texttt{ThetaEvent} embedding space and query retrieval via t-SNE projection. (Left) Embeddings for visual features, auditory features, captions, and transcriptions in one event. The semantic summary ($\mathbf{S}_\theta$, green star) is central. (Right) Text query embeddings (``dorm,'' ``football'' - stars) retrieve closest caption/transcription embeddings (crosses), linking queries to specific multimodal segments (corresponding frames and text shown).}
      \label{fig:consolidation_embedding_viz}
\end{figure*}
\subsection{Main Results and Ablation Studies}

\noindent\textbf{HippoMM yields superior accuracy and efficiency.} Table~\ref{tab:method_comparison} presents our results on the HippoVlog benchmark, designed to evaluate long-form, cross-modal associative recall.
HippoMM achieves state-of-the-art \textbf{78.2\%} accuracy, outperforming strong baselines like Ola~\citep{liu2025ola} (74.9\%) and VideoLLaMA 3~\citep{zhang2025videollama3} (73.0\%).
Notably, we implemented a parameter-free cognitive baseline, \textbf{Simple-Assoc-Memory (SAM)}, which uses classic Hebbian auto-association on the same episodic features as HippoMM. Its poor performance (30.3\%) confirms that a naive cognitive mapping is insufficient, validating the necessity of our structured architectural components.
Furthermore, HippoMM demonstrates superior efficiency: 46.2\% reduction in processing time and $5\times$ faster response time than VideoRAG~\citep{ren2024videorag}, validating that biological structure enables both accuracy and efficiency gains.

Ablation studies in Table~\ref{tab:method_comparison} validate each component.
Removing the Detailed Recall path (\textbf{HippoMM w/o DR}) causes accuracy to drop to 61.2\%, confirming its necessity for fine-grained perceptual reasoning.
Conversely, removing the Fast Retrieval path (\textbf{HippoMM w/o FR}) maintains accuracy (74.6\%) but increases average response time to 19.54s, demonstrating the semantic index's critical role in efficiency.
A pure \textbf{Embedding-Only Retrieval (EOR-only)} baseline still achieves 71.1\%, proving the inherent strength of the underlying memory structure independent of LLM reasoning.
Finally, replacing GPT-4o with the smaller \textbf{Qwen2.5-14B} maintains strong performance (70.8\%), confirming that our cognitive architecture rather than model scale drives effectiveness.
\subsection{Evaluation of Generalization and Core Capabilities}

\noindent\textbf{Episodic segmentation enables better temporal understanding.} We evaluated HippoMM through a comprehensive protocol spanning event representation and temporal understanding (Table~\ref{tab:evaluation_scores}). For event representation, we measure Content Relevance (R), Semantic Similarity (S), and Action Alignment (A) using Charades~\cite{sigurdsson2016hollywood} with GPT-4o-based evaluation. For temporal understanding, we employ the Needle-in-a-Haystack (NQA) task from MLVU~\cite{zhou2025mlvu}, which requires locating specific information within videos exceeding five minutes. HippoMM achieves 73.1\% on NQA, a 46\% improvement over VideoLlama~2, while also leading in Semantic Similarity and Action Alignment. Qwen2.5-Omni performs well on shorter videos but did not yield NQA results under our evaluation setup. These results validate that hippocampus-inspired episodic segmentation enables superior temporal navigation. On Video-MME~\cite{chen2025videomme} (61.8\%) and LongVideoBench~\cite{wu2024longvideobench} (55.2\%), HippoMM shows competitive performance; full cross-benchmark comparison is in the Appendix (Table~\ref{tab:generalization}).

\begin{table}[htbp]
\centering
\caption{Evaluation scores across core capabilities. \textbf{R}: Relevance, \textbf{S}: Similarity, \textbf{A}: Alignment (1-5 scale). \textbf{NQA}: Needle QA (\%). \textbf{Overall}: Mean normalized score (\%). GPT-4o included as a strong reference baseline. Best results in \textbf{bold}, second best \underline{underlined}.}
\label{tab:evaluation_scores}
\resizebox{\linewidth}{!}{
\begin{tabular}{lccccc}
\toprule
\textbf{Method} & \textbf{R} & \textbf{S} & \textbf{A} & \textbf{NQA (\%)} & \textbf{Overall (\%)} \\
\midrule
Qwen2.5-Omni & \textbf{3.39} & \underline{3.14} & \underline{2.76} & -- & \underline{62.0} \\
VideoLlama 2 & 2.32 & 2.17 & 1.59 & \underline{49.9} & 42.9 \\
\rowcolor{blue!20} \textbf{HippoMM} & \underline{3.28} & \textbf{3.30} & \textbf{3.14} & \textbf{73.1} & \textbf{66.9} \\
\rowcolor{gray!20} GPT-4o & 3.62 & 3.44 & 3.02 & 64.8 & 66.6 \\
\bottomrule
\end{tabular}
}
\end{table}

\subsection{Analysis of HippoMM Mechanisms}
\label{ssec:mechanism_analysis}

\noindent\textbf{Consolidation enables cross-modal pattern completion.}\label{sssec:consolidation_embedding_viz} Figure~\ref{fig:consolidation_embedding_viz} visualizes how memory consolidation creates unified \texttt{ThetaEvent} representations. The semantic summary ($\mathbf{S}_\theta$) occupies a central position among modality-specific clusters, integrating information across perceptual inputs. In retrieval, text queries like ``dorm'' successfully retrieve semantically relevant content (``stay'' transcription with residential visuals) despite lexical mismatch, while ``football'' precisely retrieves ``first football game'' mentions with field visuals. This demonstrates how consolidated structures enable both abstract summarization and specific detail retrieval from partial cross-modal cues.

\begin{figure}[H]
  \centering
  \includegraphics[width=\linewidth]{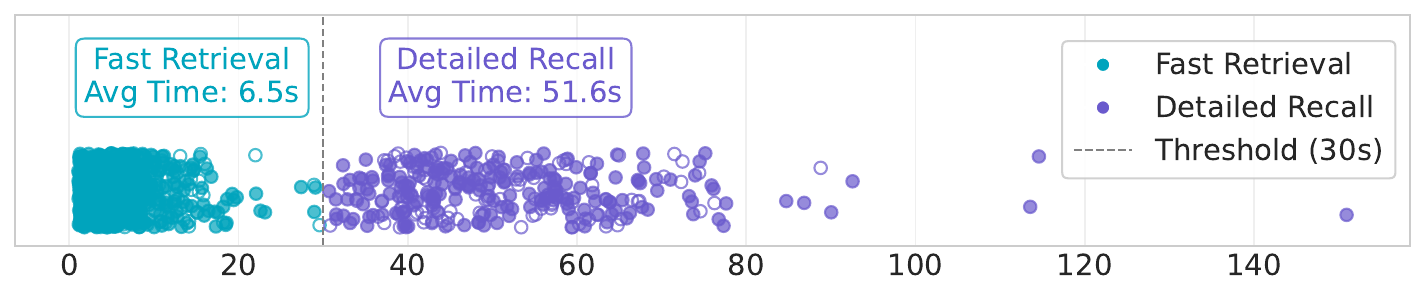}
\caption{Retrieval pathway analysis with x-axis showing response time. Blue markers ($<$30s) and purple markers ($\geq$30s) represent fast and detailed pathways respectively; filled markers indicate correct outcomes, open markers incorrect. Dashed line marks the 30s threshold.}
  \label{fig:pathway_usage}
\end{figure}

\noindent\textbf{Hierarchical Memory Retrieval reduces computational demands.} Figure~\ref{fig:pathway_usage} analyzes our hierarchical memory retrieval dynamics. Fast retrieval successfully handles 69.1\% of queries with 6.5s average response time, while detailed retrieval processes complex queries requiring fine-grained perceptual information in 51.6s. This 8x speedup for the majority of queries, achieved without sacrificing accuracy, confirms that biological hierarchical structure yields computational efficiency through intelligent query routing.
Our GPT-4o query classifier achieves 94.1\% routing accuracy; even misrouted queries achieve 52.5\% accuracy, demonstrating graceful degradation.

\noindent\textbf{Episodic segmentation preserves event integrity.} We quantify segmentation quality through Event Integrity Rate (EIR), measuring the fraction of ground-truth events fully contained within single segments:
\begin{equation}
    \text{EIR} = \frac{1}{|E_{gt}|} \sum_{e \in E_{gt}} \mathbb{I}(\exists s \in S_{gen} : s \text{ fully contains } e)
    \label{eq:eir}
\end{equation}
where $E_{gt}$ denotes ground-truth events, $S_{gen}$ denotes our generated segments, and $\mathbb{I}(\cdot)$ is the indicator function.
On Charades \cite{sigurdsson2016hollywood} with ground-truth action annotations, our episodic segmentation achieves 61.6\% EIR versus 40.2\% for fixed 15-second chunks.
This improvement demonstrates superior preservation of semantic event structure, crucial for memory formation and query accuracy.

\noindent\textbf{Consolidation provides noise resilience.} We evaluated HippoMM's robustness by introducing varying levels of perceptual noise to the HippoVlog dataset. For visual noise, we applied \texttt{ffmpeg}'s noise filter with \texttt{alls} parameters of 10, 25, and 50 for low, medium, and high levels respectively. For auditory noise, we mixed white noise at 0.1, 0.3, and 0.5 volume ratios. Under visual noise, overall accuracy decreased from the clean baseline of 78.2\% to 75.7\% (low), 75.0\% (medium), and 74.4\% (high); under auditory noise, accuracy declined to 77.5\%, 77.0\%, and 74.7\% respectively. Even under severe conditions, maximum degradation remains below 4 percentage points. This graceful degradation validates that consolidated multimodal representations provide effective redundancy against perceptual corruption, leveraging both pre-trained encoder robustness and the error-correcting properties of cross-modal consolidation.

\noindent\textbf{Hyperparameter Robustness.} We validated that our key hyperparameters ($\tau$ and $\tau_v$) are empirically optimal. The retrieval threshold ($\tau=0.75$) provides a clear speed-accuracy trade-off: lower values needlessly increase ART, while higher values cause accuracy to drop as detailed recall is bypassed. Similarly, the visual segmentation threshold ($\tau_v=0.65$) is at an optimal point: lower values over-segment and shatter events, while higher values under-segment and incorrectly merge them, with both cases degrading final accuracy. Our default values are thus robust and empirically justified.

\section{Conclusion}
\label{sec:conclusion}
HippoMM implements hippocampal principles of pattern separation, associative recall, and memory consolidation for multimodal understanding, achieving 78.2\% accuracy on HippoVlog while maintaining 5× faster response than retrieval-augmented baselines. Ablations confirm each component's necessity, while hierarchical retrieval balances efficiency with quality. This work demonstrates that cognitive architectures grounded in neuroscience provide principled foundations for multimodal computing systems, suggesting biological memory mechanisms offer promising pathways toward human-like understanding in artificial systems.

\section*{Acknowledgments}
This work was supported in part by NSF CNS-2112562 and ARO W911NF-23-2-0224.
{
    \small
    \bibliographystyle{ieeenat_fullname}
    \bibliography{main}
}

\onecolumn
\appendix
\section*{Appendix}
\section{Key Notation}
\label{sec:appendix_notation}

\begin{table}[h]
\caption{Key notation used throughout the paper}
\centering
\small
\begin{tabular}{ll}
\toprule
\textbf{Symbol} & \textbf{Description} \\
\midrule
$X = \{x_t\}$ & Continuous multimodal data stream \\
$\mathcal{S}_t$ & Segmentation trigger indicator at time $t$ \\
$\mathbbm{1}[\cdot]$ & Indicator function (returns 1 if true, 0 otherwise) \\
$m_i$ & $i$-th \texttt{ShortTermMemory} object \\
$K$ & Set of indices for consolidated memories \\
$k \in K$ & Index in the consolidated set \\
$m_k$ & Consolidated \texttt{ShortTermMemory} object \\
$\theta_k$ & $k$-th ThetaEvent object \\
$\mathcal{M}$ & Collection of consolidated \texttt{ShortTermMemory} objects \\
$\Theta$ & Collection of ThetaEvent objects \\
$\mathbf{v}_i, \mathbf{v}_k$ & Representative embedding for segment $i$ or $k$ \\
$\mathbf{E}_i$ & Sequence of cross-modal embeddings for segment $i$ \\
$\mathbf{T}_i$ & Aggregated textual outputs for segment $i$ \\
$\mathbf{C}_i$ & Pointers to raw audiovisual data for segment $i$ \\
$\mathbf{S}_{\theta_k}$ & Semantic summary of ThetaEvent $k$ \\
$\mathcal{E}_c$ & Cross-modal embedding model \\
$\mathcal{T}_v$ & Vision-language model for visual descriptions \\
$\mathcal{T}_a$ & Speech recognition model \\
$\phi_{\text{LLM}}$ & LLM for semantic summary generation \\
$\gamma$ & Consolidation similarity threshold \\
$\tau$ & Confidence threshold for retrieval switching \\
$\delta$ & Temporal window expansion parameter \\
$\cos(\cdot, \cdot)$ & Cosine similarity function \\
$\text{sim}(\cdot, \cdot)$ & Similarity function (cosine similarity) \\
$\text{TopK}(\cdot, k)$ & Function returning $k$ highest-scoring elements \\
$\text{FormattedContext}(\cdot)$ & Function to structure multimodal content \\
$\text{Extract}_{\text{target}}(\cdot)$ & Function to extract target modality content \\
$(t_{s,i}, t_{e,i})$ & Start and end times of segment $i$ \\
$[t_{s,i}, t_{e,i}]$ & Alternative notation for segment boundaries \\
$\Phi_{\text{fast}}$ & Fast retrieval function \\
$\Psi_{\text{detailed}}$ & Detailed recall function \\
$\Gamma$ & Adaptive reasoning module \\
$\mathcal{C}_{\text{context}}$ & Context information for reasoning \\
\bottomrule
\end{tabular}
\end{table}

\section{HippoVlog Construction Details}
\label{sec:appendix_construction}

The HippoVlog dataset was created through a systematic pipeline designed to capture rich multimodal information from daily vlogs and generate challenging memory-based questions. The process follows five main stages: video selection, multimodal data sampling, ground truth generation, question generation, and curation.

\subsection{Video Selection and Preprocessing}
\label{subsec:appendix_video_selection}

\begin{itemize}
    \item \textbf{Source Selection:} 25 daily vlogs from YouTube (Creative Commons licensed), with average duration of 27.3 minutes, totaling approximately 682 minutes of content.
    \item \textbf{Quality Criteria:} Videos were chosen based on clear audio and visual quality, natural episodic structure typical of daily life, and diverse activities (cooking, shopping, travel).
    \item \textbf{Format Standardization:} All source videos were preprocessed to a standard format (720p resolution at 30fps) with normalized audio levels.
\end{itemize}

\subsection{Multimodal Data Sampling}
\label{subsec:appendix_sampling}

To create manageable and representative data points from the full vlogs, we implemented a robust sampling strategy for both audio and video modalities:

\begin{itemize}
    \item \textbf{Audio Segment Extraction:}
        \begin{itemize}
            \item 10 non-overlapping audio segments (each 10 seconds long) were randomly selected per video.
            \item Sampling avoided the initial and final 5\% of videos to exclude typical intros/outros.
            \item Silent segments were discarded and resampled.
        \end{itemize}
    \item \textbf{Video Segment Extraction:}
        \begin{itemize}
            \item Corresponding 10-second video segments were extracted for each valid audio segment.
            \item Checks prevented significant overlap between sampled segments.
            \item Metadata about the sampled segments was preserved for future reference.
        \end{itemize}
\end{itemize}

\subsection{Segment-Level Ground Truth Generation}
\label{subsec:appendix_segment_captioning}

Detailed descriptions combining audio and visual information were generated for each sampled 10-second video segment:

\begin{itemize}
    \item \textbf{Model Used:} Qwen2.5-Omni-7B.
    \item \textbf{Process:} The model processed each video segment to generate comprehensive captions.
    \item \textbf{Structured Output:} For each segment, a structured caption containing:
        \begin{itemize}
            \item Detailed audiovisual description
            \item List of distinct visual elements
            \item Transcription of speech or description of other audio elements
            \item All content maintained anonymity (using ``the person'', ``the speaker'')
        \end{itemize}
\end{itemize}

\subsection{Question Generation}
\label{subsec:appendix_question_generation}

Multiple-choice questions were generated across four categories, with visual questions specifically based on individual frames extracted from video segments:

\begin{itemize}
    \item \textbf{Models Used:} GPT-4o.
    \item \textbf{Input:} For visual questions, the model received frames and corresponding video context.
    \item \textbf{Design Principles:} Questions were designed to:
        \begin{itemize}
            \item Focus on specific, distinctive details
            \item Be potentially answerable if the video had been previously seen
            \item Maintain anonymity
            \item Have exactly one correct answer among four plausible options
        \end{itemize}
\end{itemize}

\subsection{Question Curation and Finalization}
\label{subsec:appendix_curation}

The automatically generated questions underwent further curation:

\begin{itemize}
    \item \textbf{Initial Pool:} Approximately 1,500 candidate questions generated across all types.
    \item \textbf{Quality Control:}
        \begin{itemize}
            \item Manual validation by human.
            \item Questions that were ambiguous, subjective, or had multiple plausible answers were removed or refined.
        \end{itemize}
    \item \textbf{Final Dataset:} 1,000 high-quality questions, balanced across four categories (250 per category).
    \item \textbf{Answer Generation:} Each question includes one correct answer and three plausible but verifiably incorrect distractors, balanced for length and style.
\end{itemize}

\subsection{Question Categories and Examples}
\label{subsec:appendix_categories}

Each of the four question categories in HippoVlog targets a specific memory function:

\subsubsection{Cross-Modal Binding (T\textsubscript{V$\times$A})}
Questions requiring integration of information across visual and auditory modalities.

\begin{tcolorbox}[title=Example: Cross-Modal Binding]
\textbf{Question:} ``When the main character is sitting in the car wearing a black jacket and white scarf, what does she audibly express?''

\textbf{Options:}
\begin{itemize}
    \item A: She says she feels excited about the drive.
    \item B: She mentions being scared and says it feels weird.
    \item C: She talks about the weather being too cold.
    \item D: She expresses confidence about her driving skills.
\end{itemize}
\textbf{Explanation:} In the video, as the main character sits in the car wearing a black jacket and white scarf, she audibly expresses her nervousness by saying `okay, hey, I'm scared, oh my gosh this feels weird.'

\end{tcolorbox}

\subsubsection{Auditory-Focused Retrieval (T\textsubscript{A})}
Questions focusing on auditory information, including speech content, background sounds, and music.

\begin{tcolorbox}[title=Example: Auditory-Focused Retrieval]
\textbf{Question:} ``What did the main character mention about the purpose of the meeting?''

\textbf{Options:}
\begin{itemize}
    \item A: To finalize a business deal
    \item B: To discuss the upcoming project
    \item C: To plan a surprise party
    \item D: To talk about a vacation
\end{itemize}
\textbf{Explanation:} The main character specifically mentions the `upcoming project' as the purpose of the meeting, indicating a discussion centered around future plans and tasks.
\end{tcolorbox}

\subsubsection{Visual-Focused Retrieval (T\textsubscript{V})}
Questions targeting visual details that require careful observation of scene elements.

\begin{tcolorbox}[title=Example: Visual-Focused Retrieval]
\textbf{Question:} ``What distinctive pattern is visible on the roof of the building?''

\textbf{Options:}
\begin{itemize}
    \item A: Checkerboard
    \item B: Chevron
    \item C: Striped
    \item D: Polka dot
\end{itemize}

\textbf{Explanation:} The roof of the building in the image has a distinctive chevron pattern.
\end{tcolorbox}

\begin{tcolorbox}[title=Example: Visual-Focused Retrieval]
\textbf{Question:} ``What is the main character wearing on her head?''

\textbf{Options:}
\begin{itemize}
    \item A: A baseball cap
    \item B: A knitted hat
    \item C: A colorful headscarf
    \item D: A headband
\end{itemize}

\textbf{Explanation:} The main character is wearing a colorful headscarf, as seen in the image.
\end{tcolorbox}

\subsubsection{Semantic/Temporal Reasoning (T\textsubscript{S})}
Questions requiring integration of information across temporal segments and abstract understanding.

\begin{tcolorbox}[title=Example: Semantic/Temporal Reasoning]
\textbf{Question:} ``What was one significant challenge the main character faced during the experiment?''

\textbf{Options:}
\begin{itemize}
    \item A: Finding time to eat breakfast
    \item B: Dealing with brain fog
    \item C: Avoiding social media
    \item D: Lack of exercise options
\end{itemize}

\textbf{Explanation:} Later in the vlog, after the experiment has been underway for several hours, the main character states they are ``dealing with a lot of brain fog,'' which is affecting their focus on the task.
\end{tcolorbox}

\section{HippoMM Implementation Details}

\label{app:implementation}

This appendix provides the specific implementation choices, models, and parameters used in the HippoMM system, complementing the conceptual description in the Methodology section.

\paragraph{System Configuration}

All experiments reported were conducted on a single NVIDIA L40S GPU.

\subsection{Memory Formation Implementation}

\paragraph{Temporal Pattern Separation}

The process segments continuous video streams based on perceptual changes:

\begin{itemize}
    \item \textbf{Input Sampling:} Raw video frames are adaptively sampled at a rate between 1 and 10 frames per second (fps). Raw audio is sampled at 16 kHz PCM.
    \item \textbf{Visual Change Detection:} A boundary is triggered ($\mathcal{S}_t=1$) if the visual dissimilarity between consecutive frames, measured as $d_v(F_t, F_{t-1}) = 1 - \text{SSIM}(F_t, F_{t-1})$ (Eq.~\ref{eq:ssim_distance}), exceeds a threshold $\tau_v = 0.65$.
    \item \textbf{Auditory Change Detection:} A boundary is triggered ($\mathcal{S}_t=1$) if the negative log energy of the raw audio signal, $d_a(a_t)$ (Eq.~\ref{eq:audio_energy}), exceeds a threshold $\tau_a = 40$. This corresponds to the signal amplitude dropping below a threshold (approx. -40 dB), indicating potential silence or a significant break.
    \item \textbf{Segment Duration:} The resulting segments are constrained to be between $t_{min} = 5$ seconds and $t_{max} = 10$ seconds long. Shorter segments following boundary detection are merged with preceding ones, and longer intervals between boundaries are split to adhere to this constraint, defining the temporal windows $[t_{start, i}, t_{end, i}]$.
\end{itemize}

\paragraph{Perceptual Encoding}

Each segment $i$ (defined by $[t_{start, i}, t_{end, i}]$) is processed to extract multimodal features:

\begin{itemize}
    \item \textbf{Cross-Modal Embeddings:} We use ImageBind (v1.0) as the joint embedding model ($\mathcal{E}_c$) to generate sequences of shared 1024-dimensional embeddings ($\mathbf{E}_i$) from visual frames and audio snippets within the segment.
    \item \textbf{Audio Transcription:} Whisper (v3 medium) is employed as the speech recognition model ($\mathcal{T}_a$) to obtain time-aligned transcriptions from the segment's audio.
    \item \textbf{Visual Description:} Qwen2.5-VL is used as the vision-language model ($\mathcal{T}_v$) to generate textual descriptions of the visual content (maximum length 200 tokens).
    \item \textbf{\texttt{ShortTermMemory} Object ($m_i$):} These processed elements are aggregated and stored in a \texttt{ShortTermMemory} object $m_i$, as defined in Eq.~(\ref{eq:short_term_memory}). It encapsulates: pointers to the raw audiovisual snippets ($\mathbf{C}_i$), the sequence of ImageBind embeddings ($\mathbf{E}_i$), the aggregated Whisper transcriptions and Qwen-VL visual descriptions ($\mathbf{T}_i$), segment start/end timestamps ($t_{s,i}, t_{e,i}$), and related metadata.
\end{itemize}

\paragraph{Memory Consolidation}

Redundancy is reduced by filtering similar consecutive segments:

\begin{itemize}
    \item \textbf{Similarity Calculation:} A representative cross-modal embedding ($\mathbf{v}_i$) is calculated for each segment $m_i$ by averaging its sequence of ImageBind embeddings ($\mathbf{E}_i$).
    \item \textbf{Filtering Criterion:} A segment $m_i$ is retained in the consolidated set $K$ only if the cosine similarity between its representative embedding $\mathbf{v}_i$ and the representative embedding $\mathbf{v}_j$ of the most recent previously retained segment $m_j$ (where $j \in K, j<i$) is below a predefined threshold $\gamma = 0.85$, as per Eq.~\ref{eq:consolidation_filtering}. Segments not meeting this criterion are discarded (or merged). The output is the sequence of consolidated objects $\{m_k \mid k \in K\}$.
\end{itemize}

\paragraph{Semantic Replay}

Abstracted representations are generated for long-term storage:

\begin{itemize}
    \item \textbf{Summarization Model:} The Qwen2.5-VL large vision-language model ($\phi_{\text{LLM}}$) processes formatted multimodal context derived from the consolidated segment $m_k$ (including key visual information $\mathbf{I}_{\text{visual}}(m_k)$ and auditory information $\mathbf{A}_{\text{audio}}(m_k)$). It generates a concise textual summary ($\mathbf{S}_{\theta_k}$) using a temperature of 0.7 and a maximum length of 300 tokens (see Table~\ref{tab:prompt} for prompt details), as per Eq.~\ref{eq:semantic_summary_revised}.
    \item \textbf{\texttt{ThetaEvent} Object ($\theta_k$):} This object constitutes the long-term memory entry, storing the elements defined in Eq.~\ref{eq:theta_event}: the representative cross-modal embedding ($\mathbf{v}_k$), the generated textual summary ($\mathbf{S}_{\theta_k}$), references to the key visual ($\mathbf{I}_{\theta_k}$) and auditory ($\mathbf{A}_{\theta_k}$) context used, and the segment's temporal information ($\mathbf{T}_{\theta_k}$) which links back to the detailed \texttt{ShortTermMemory} object $m_k$. The collection $\{\theta_k\}$ forms the abstract long-term memory store.
\end{itemize}

\subsection{Memory Retrieval Implementation}

\paragraph{Query Analysis}

User queries are initially processed to determine the retrieval strategy:

\begin{itemize}
    \item \textbf{Classifier Model:} GPT-4o is used as the query classifier ($\mathcal{Q}_{\text{type}}$) to classify the incoming query $q$ based on its primary modality focus (Visual, Auditory, Cross-modal) or if it targets high-level semantic/gist information (Summary). This uses a 3-shot prompt with examples (see Table~\ref{tab:query_classification}).
\end{itemize}

\paragraph{Hierarchical Retrieval}

Retrieval follows a confidence-gated two-pathway approach (Eq.~\ref{eq:hierarchical_retrieval}):

\begin{itemize}
    \item \textbf{Fast Retrieval ($\Phi_{\text{fast}}$):} Retrieval is first attempted using the abstract textual summaries ($\mathbf{S}_{\theta_k}$) stored in the collection of \texttt{ThetaEvent} objects ($\mathbf{\Theta} = \{\theta_k\}$). Confidence scores are determined by internal LLM calibration (Qwen2.5-VL, see Table~\ref{tab:direct_answer}) on a scale of 0-1.
    \item \textbf{Detailed Recall ($\Psi_{\text{detailed}}$):} If the confidence score from Fast Retrieval falls below a threshold $\tau = 0.75$, retrieval escalates to the Detailed Recall pathway. This pathway searches the consolidated \texttt{ShortTermMemory} objects ($\mathcal{M} = \{m_k \mid k \in K\}$), guided by the query type determined by $\mathcal{Q}_{\text{type}}(q)$.
\end{itemize}

\paragraph{Detailed Recall Pathways}

The specific operations within Detailed Recall ($\Psi_{\text{detailed}}$) are guided by the classified query type:

\begin{itemize}
    \item \textbf{Query Embedding:} The query $q$ is converted into an ImageBind embedding ($q_{embed}$) for use in similarity-based search.
    \item \textbf{Similarity Search (Feature/Semantic):} For single-modality or cue-based retrieval (Visual or Auditory query type), cosine similarity is used to find the top-$k$ \texttt{ShortTermMemory} segments ($m_k$) whose representative embeddings ($\mathbf{v}_k$) are most similar to $q_{embed}$ (Feature Search), with $k = 5$. Text-based search over the aggregated descriptions/transcriptions ($\mathbf{T}_k$) serves as a fallback or alternative (Semantic Search). See Tables~\ref{tab:frame_selection} and \ref{tab:audio_selection} for related selection prompts.
    \item \textbf{Cross-Modal Association:} For Cross-modal query types, retrieval uses temporal co-occurrence. First, the top-$k$ seed segments $\mathbf{S}_{\text{query}}$ most similar to $q_{embed}$ (based on $\mathbf{v}_k$) are identified (Eq.~\ref{eq:cross_modal_query_retrieve_revised}, with $k=5$). Second, expanded temporal windows ($\mathbf{W}$) are defined around these seeds using a temporal buffer parameter $\delta = 2$ seconds (Eq.~\ref{eq:cross_modal_window}). Third, target modality information ($\mathbf{S}_{\text{target}}$) is extracted from any segment $m_j$ whose time interval overlaps with any window in $\mathbf{W}$ (Eq.~\ref{eq:cross_modal_target_retrieve}).
    \item \textbf{Fine-grained Analysis:} If necessary for answering the query, specific content within the retrieved \texttt{ShortTermMemory} segments is analyzed further using Qwen2.5-VL (for visual/multimodal questions) or Whisper (for detailed audio content examination). This often occurs during the final synthesis stage.
\end{itemize}

\paragraph{Adaptive Reasoning}

The final answer is synthesized from the retrieved information:

\begin{itemize}
    \item \textbf{Synthesis Model:} Qwen2.5-VL is typically used as the reasoning module ($\Gamma$) to generate the final answer ($a$). It takes the original query ($q$), the retrieved evidence ($r_{\text{retrieved}}$), and relevant extracted context ($\mathcal{E}_{\text{context}}$) as input (Eq.~\ref{eq:adaptive_reasoning}, see Table~\ref{tab:final_answer} for prompt). For complex reasoning tasks requiring deeper integration or handling potential conflicts (e.g., comparing Fast Retrieval and Detailed Recall outputs), GPT-4o may be used with a 4-shot prompting strategy and temperature of 0.2 (see Table~\ref{tab:reflection} for a related prompt).
\end{itemize}

\section{Cross-Benchmark Generalization}
\label{sec:appendix_generalization}

\begin{table}[t]
\centering
\caption{Cross-benchmark generalization comparison. Video-MME and LongVideoBench test general comprehension; MLVU-NQA tests precise temporal retrieval, where HippoMM's episodic memory excels. Best results in \textbf{bold}.}
\label{tab:generalization}
\small
\begin{tabular}{lccc}
\toprule
\textbf{Method} & \textbf{Video-MME} & \textbf{LongVideoBench} & \textbf{MLVU-NQA} \\
\midrule
InternVL2~\citep{chen2024far} & 61.2\% & 59.3\% & 48.3\% \\
VideoLLaMA 3~\citep{zhang2025videollama3} & \textbf{66.2\%} & \textbf{59.8\%} & 68.3\% \\
GPT-4o~\citep{hurst2024gpt} & 71.9\% & 66.7\% & 64.8\% \\
\rowcolor{blue!10} \textbf{HippoMM (Ours)} & 61.8\% & 55.2\% & \textbf{73.1\%} \\
\bottomrule
\end{tabular}
\end{table}

\section{Additional Experimental Details}

\subsection{Ablation Studies Parameter Settings}

Acronyms refer to components described in the Methodology section: FR (Fast Retrieval, $\Phi_{\text{fast}}$), DR (Detailed Recall, $\Psi_{\text{detailed}}$), AR (Adaptive Reasoning, $\Gamma$).
\begin{itemize}
    \item \textbf{HippoMM w/o DR, AR:} Uses Fast Retrieval ($\Phi_{\text{fast}}$) only, with direct answer generation from summaries (no Adaptive Reasoning $\Gamma$).
    \item \textbf{HippoMM w/o FR, AR:} Uses Detailed Recall ($\Psi_{\text{detailed}}$) only, with direct answer generation from retrieved details (no Adaptive Reasoning $\Gamma$).
    \item \textbf{HippoMM w/o AR:} Uses both retrieval paths ($\Phi_{\text{fast}}$, $\Psi_{\text{detailed}}$) as per the hierarchical logic, but the final output is based on direct extraction or simple aggregation rather than the full synthesis step ($\Gamma$).
\end{itemize}

\subsection{Performance Metrics}

\begin{itemize}
\item \textbf{Processing Time (PT):} Total time to execute the Memory Formation phase on raw videos.
\item \textbf{Average Response Time (ART):} Average time to execute the Memory Retrieval phase and generate an answer $a$ for a query $q$.
\item \textbf{Accuracy Metrics:} Based on strict binary match with human-annotated ground truth answers.
\end{itemize}

\subsection{Generalization Evaluation}

\begin{itemize}
    \item \textbf{Charades Dataset:} 100 randomly selected video clips.
    \item \textbf{MLVU Needle-in-a-Haystack:} 415 queries across 150 long-form videos (5+ minutes each).
    \item \textbf{GPT-4o Evaluation:} Used as judge for Content Relevance, Semantic Similarity, and Action Alignment with a 5-point Likert scale.
    \item \textbf{Overall Score Calculation:} Arithmetic mean after linearly normalizing 1-5 scales to 0-100\%.
\end{itemize}

\subsection{Additional Implementation Parameters}

\begin{itemize}
    \item \textbf{Key Frame Filtering Threshold:} Within Perceptual Encoding, a visual similarity threshold (SSIM-based, value 0.3) is used to filter visually redundant frames before description generation to reduce processing load. (Note: This is distinct from the consolidation threshold $\gamma$).
    \item \textbf{Visual Similarity Fallback Threshold (Detailed Recall):} When performing Feature Search in Detailed Recall for visual queries, if the max similarity score is below 0.4, the system may fall back to Semantic Search using text captions.
    \item \textbf{Direct Answer Confidence Threshold ($\tau$):} The threshold used in Hierarchical Retrieval (Eq.~\ref{eq:hierarchical_retrieval}) to switch from Fast Retrieval to Detailed Recall is $\tau = 0.75$.
    \item \textbf{Context Length (LLM):} A maximum context length of 120,000 tokens is used for GPT-4o when invoked for complex reasoning (Adaptive Reasoning $\Gamma$).
    \item \textbf{Multiple Worker Processing:} Multiprocessing with a pool size equal to the number of available CPU cores (up to a maximum of 8) is used for parallel frame processing during Perceptual Encoding.
\end{itemize}

\begin{table*}[!ht]
\caption{\label{tab:prompt} Semantic Replay Summarization Prompt ($\phi_{\text{LLM}}$)}
\begin{tcolorbox}
{\bf Task Description}
Please provide a concise one sentence summary of this event based on the video frames descriptions and audio transcription. What is happening in this event?

{\bf Input Format}
\begin{itemize}
\item Image descriptions: [Concatenated frame captions from key visual context $\mathbf{I}_{\text{visual}}(m_k)$]
\item Audio transcription: [Concatenated transcribed text from key auditory context $\mathbf{A}_{\text{audio}}(m_k)$]
\end{itemize}

{\bf Output Requirements}
\begin{itemize}
\item One sentence summary ($\mathbf{S}_{\theta_k}$) capturing the integrated multimodal essence of the event
\item Focus on key actions, subjects, and setting
\item Maintain factual accuracy based on provided descriptions
\item Avoid introducing information not supported by the input
\end{itemize}
\end{tcolorbox}
\end{table*}

\begin{table*}[ht]
\caption{\label{tab:query_classification} Query Type Classification Prompt ($\mathcal{Q}_{\text{type}}$)}
\begin{tcolorbox}
{\bf Task Description}
Classify this question into one of these categories based on what type of information is needed to answer it:

{\bf Classification Categories}
\begin{itemize}
\item VIDEO - Questions specifically about visual elements, appearances, or actions that need frame-by-frame analysis, e.g., ``what is the main character holding?'' (Corresponds to Visual $V$)
\item AUDIO - Questions about sounds, speech, or audio content that need audio analysis, e.g., ``what does the main character say/mention?'' (Corresponds to Auditory $A$)
\item VIDEO+AUDIO - Questions that require associating both visual and audio information, e.g., ``what is the main character doing while saying/mentioning something?'' (Corresponds to Cross-modal $V \times A$)
\item SUMMARY - Questions that focus on the overall content or gist of the video, e.g., ``what is the main character doing?'' (Corresponds to Semantic/Gist)
\end{itemize}

{\bf Guidelines}
\begin{itemize}
\item If question is about visual details, appearances, or actions, classify as VIDEO
\item If question is about sounds, speech, or audio content, classify as AUDIO
\item If question requires both visual and audio information, classify as VIDEO+AUDIO
\item If question is about the overall content of the video, classify as SUMMARY
\end{itemize}

{\bf Input}
Question: [user query $q$]

{\bf Output Format}
Return ONLY one of these exact words: VIDEO, AUDIO, VIDEO+AUDIO, SUMMARY
\end{tcolorbox}
\end{table*}

\begin{table*}[ht]
\caption{\label{tab:direct_answer} Fast Retrieval Attempt Prompt (Part of $\Phi_{\text{fast}}$)}
\begin{tcolorbox}
{\bf Task Description}
Given the following question and relevant event summaries ($\mathbf{S}_{\theta_k}$ from $\mathbf{\Theta}$), analyze whether the question can be answered directly or needs specific analysis (triggering Detailed Recall $\Psi_{\text{detailed}}$).

{\bf Input}
Question: [user query $q$]

Event Summaries:
[List of relevant event summaries $\mathbf{S}_{\theta_k}$ retrieved based on $q$]

{\bf Guidelines for analysis}
\begin{itemize}
\item General questions about overall video content (likely classified as SUMMARY) should be attempted directly from summaries.
\item Questions about specific visual details (VIDEO), sounds/speech (AUDIO), or combined elements (VIDEO+AUDIO) often require specific analysis beyond summaries.
\item Estimate confidence based on whether summaries contain sufficient information.
\end{itemize}

{\bf Output Format}
Must be in one of these two structures:

1. If answerable from summaries:
\begin{verbatim}
ANSWER: <your detailed answer based ONLY on summaries,
if multiple choice, output one letter>
CONFIDENCE: <score between 0.0-1.0, e.g., 0.7 or higher if confident>
\end{verbatim}

2. If requiring specific analysis (summaries insufficient):
\begin{verbatim}
ANSWER: NONE
CONFIDENCE: <low score, e.g., 0.0 or < 0.7>
\end{verbatim}
(Note: This confidence score is compared against threshold $\tau$)
\end{tcolorbox}
\end{table*}

\begin{table*}[ht]
\caption{\label{tab:frame_selection} Visual Segment Selection Prompt (Part of $\Psi_{\text{detailed}}$ - Semantic Search)}
\begin{tcolorbox}
{\bf Task Description}
Given a question (potentially classified as VIDEO or VIDEO+AUDIO), descriptions from consolidated memory segments ($\mathbf{T}_k$), and an optional element to search for, identify at most 5 relevant frame indices (within segments $m_k$) for answering the question.

{\bf Input}
Question: [user query $q$]
Element to search for: [search query derived from $q$]

Frame descriptions (from relevant $\mathbf{T}_k$):
[Numbered list of frame descriptions/captions]

{\bf Instructions}
\begin{itemize}
\item Return ONLY numbers separated by commas (e.g., ``0,3,5,8,12'') corresponding to indices within the provided list.
\item Return at most 5 indices.
\item Do not include any other text, explanations, or spaces.
\item If fewer than 5 frames are relevant, return fewer indices.
\end{itemize}

{\bf Example good responses}
``0,3,5,8,12''
``1,4,7''

{\bf Output Format}
Your response (numbers only, comma-separated):
\end{tcolorbox}
\end{table*}

\begin{table*}[ht]
\caption{\label{tab:audio_selection} Audio Segment Selection Prompt (Part of $\Psi_{\text{detailed}}$ - Semantic Search)}
\begin{tcolorbox}
{\bf Task Description}
Given a question (potentially classified as AUDIO or VIDEO+AUDIO) and transcriptions with timestamps from consolidated memory segments ($\mathbf{T}_k$), identify the most relevant time frames where the answer might be found.

{\bf Input}
Question: [user query $q$]

Transcriptions (with timestamps from relevant $\mathbf{T}_k$):
[List of timestamped transcriptions]

{\bf Instructions}
\begin{itemize}
\item Analyze the transcriptions and identify segments most likely to contain the answer.
\item Return a JSON array of time frames (relative to the source video). Each frame defines a start and end time.
\begin{verbatim}
[
    {"start": START_TIME, "end": END_TIME},
    {"start": START_TIME, "end": END_TIME}
]
\end{verbatim}
\item Return at most 5 time frames.
\item Include a small buffer around identified time frames (e.g., $\pm$2 seconds, related to $\delta$).
\item If no relevant segments found, return ``[]''.
\end{itemize}

{\bf Output Format}
Your response (valid JSON only):
\end{tcolorbox}
\end{table*}

\begin{table*}[ht]
\caption{\label{tab:reflection} Adaptive Reasoning Prompt (Optional step within $\Gamma$)}
\begin{tcolorbox}
{\bf Task Description}
You potentially have two candidate answers to the same question $q$. One derived from Fast Retrieval ($\Phi_{\text{fast}}$, using summaries $\mathbf{\Theta}$) and another from Detailed Recall ($\Psi_{\text{detailed}}$, using detailed segments $\mathcal{M}$). Compare them and provide a final, reconciled answer $a$.

{\bf Input}
Question: [user query $q$]

Answer from summaries (if $\Phi_{\text{fast}}$ produced one): [direct\_answer]
Confidence from summaries: [direct\_confidence]

Answer from detailed analysis (if $\Psi_{\text{detailed}}$ was run): [detailed\_answer]

Supporting Context ($\mathcal{E}_{\text{context}}$):
Sample frame captions: [Sample relevant frame captions from $r_{\text{retrieved}}$]
Sample transcriptions: [Sample relevant transcriptions from $r_{\text{retrieved}}$]

{\bf Instructions}
\begin{itemize}
    \item Compare both answers for consistency, detail, and plausibility given the context.
    \item If they agree, prefer the more detailed/confident answer.
    \item If they disagree, evaluate which answer is better supported by the specific context (captions, transcriptions). Prioritize the answer derived from detailed analysis ($\Psi_{\text{detailed}}$) if confidence in the summary-based answer ($\Phi_{\text{fast}}$) was low or if the detailed analysis provides clearly contradictory evidence.
    \item Synthesize the best possible final answer based on this comparison.
    \item For multiple choice questions, select only one final answer option.
    \item Estimate the confidence in the final reconciled answer.
\end{itemize}

{\bf Output Format}
\begin{verbatim}
ANSWER: <reconciled final answer a>
CONFIDENCE: <final confidence score between 0.0-1.0>
REASONING: <brief explanation for the final answer choice>
\end{verbatim}
\end{tcolorbox}
\end{table*}

\begin{table*}[ht]
\caption{\label{tab:final_answer} Final Answer Synthesis Prompt ($\Gamma$)}
\begin{tcolorbox}
{\bf Task Description}
Based on the following retrieved context ($r_{\text{retrieved}}$) and extracted evidence ($\mathcal{E}_{\text{context}}$), please synthesize a final answer ($a$) to the question ($q$).

{\bf Input}
Question: [user query $q$]

Overall Video Context (e.g., relevant summaries $\mathbf{S}_{\theta_k}$ if available):
[List of relevant high-level context]

Relevant Retrieved Content (details from $\Phi_{\text{fast}}$ or $\Psi_{\text{detailed}}$):
[List of retrieved content descriptions, e.g., frame captions, transcriptions, details from $m_k$]

{\bf Instructions}
Please provide a clear and specific answer based on the integration of the overall context and the specific retrieved content. If the information is insufficient or contradictory, state that or provide the best possible reasoned answer based on the evidence.

Output should be one letter if the original question was multiple choice.

{\bf Output Format}
\begin{verbatim}
Answer: <Synthesized answer a>
\end{verbatim}
\end{tcolorbox}
\end{table*}

\end{document}